\documentstyle[aps,preprint,epsf,graphicx]{revtex}
\begin{document}
\draft
\preprint{}
\title{\textit{In-Situ} absolute phase detection of a microwave 
field via incoherent fluorescence }
\author{ George C. Cardoso$^{1}$, Prabhakar Pradhan$^{1}$, 
          Jacob Morzinski$^{2}$, and  M. S. Shahriar$^{1,2}$ }
\address{$^{1}$Department of Electrical and Computer Engineering, 
  Northwestern University,  Evanston, IL 60208 \\
  $^{2}$ Research Laboratory of Electronics, Massachusetts Institute of Technology, Cambridge, 
  MA 02139 }
\maketitle
\date{\today}
\begin{abstract}
Measuring the amplitude and the absolute phase of a monochromatic
microwave field at a specific point of space and time has many potential
applications, including precise qubit rotations and wavelength quantum
teleportation. Here we show how such a measurement can indeed be made using
resonant atomic probes, via detection of incoherent fluorescence induced by
a laser beam. This measurement is possible due to self-interference effects
between the positive and negative frequency components of the field. In  
effect, the small cluster of atoms here act as a highly localized pick-up
coil, and the fluorescence channel acts as a transmission line. 
\end{abstract}

 Measurement of the amplitude and the absolute phase of a monochromatic wave 
is challenging because in the most general condition the spatial
distribution of the field around a point is arbitrary. Therefore, one must
know the impedance of the system between the point of interest and the
detector, and ensure that there is no interference with the ambient field.
It is recently shown in the literature that the absolute phase measurement can 
be used for accurate qubit rotations [1-3] and quantum wavelength teleportation [4-6].

  Before we describe the physics behind this process, it is instructive to 
define precisely what we mean by the term ``absolute phase." Consider, for 
example, a microwave field such that the magnetic field at a position 
\textbf{\textit{R}} is given by 
\textbf{\textit{B}}$(t)= B_{0}$\textit{cos($\omega $t+$\phi )$}$\hat {x}$, where 
\textit{$\omega $ }is the frequency of the field, and \textit{$\phi $ } 
is determined simply by our choice of 
the origin of time. The absolute phase is the sum of the temporal and the 
initial phase, i.e., \textit{$\omega $t+$\phi $}. 
In order to illustrate how this phase can be observed 
directly, consider a situation where a cluster of non-interacting atoms are 
at rest at the same location. For simplicity, we assume each atom to be an 
ideal two-level system where a ground state $\vert $0$>$ is coupled to an 
excited state $\vert $1$>$ by this field \textbf{B}(t), with the atom 
initially in state $\vert $0$>$. The Hamiltonian for this interaction is:

\begin{equation}
\label{eq1}
\hat {H}=\varepsilon (\sigma _0 -\sigma _z )/2+g(t)\sigma _x ,
\end{equation}

\noindent where $g(t)=-g_{o}$\textit{cos($\omega $t+$\phi )$}, $g_{o }$is the 
Rabi frequency, $\sigma _i $ are the Pauli  matrices, and the driving frequency 
$\omega =\varepsilon $ corresponds to  resonant excitation. We consider $g_0$
to be of the form $g_0 (t)$ $=$ $g_{0M} [1-\exp (-t/\tau _{sw} )]$ with a switching time 
$\tau _{sw} $ relatively  slow compared to other time scales in the system, 
i.e. $\tau _{sw} >>$ $\omega ^{-1}$ and g$^{-1}_{0M}$. 

  As we have shown before [2,3], without the rotating wave approximation 
(RWA) and to the lowest order in $\eta \equiv $(g$_{0}$/4$\omega )$, the 
amplitudes of $\vert $0$>$ and $\vert $1$>$ at any time $t$ are as follows :
\begin{equation}
\label{eq2}
C_0 (t)=\cos ({g}'_0(t)t/2)-2\eta \Sigma \cdot \sin ({g}'_0 (t)t/2),
\end{equation}
\begin{equation}
\label{eq3}
C_1 (t)=ie^{-i(\omega {\kern 1pt}t+\phi )}[\sin ({g}'_0 (t)t/2)+2\eta 
B
\Sigma ^\ast \cdot \cos ({g}'_0 (t)t/2)],
\end{equation}
where $\Sigma \equiv (i/2)\exp [-i(2\omega t+2\phi )]$, and 
${g}'_0 (t)\equiv \frac{1}{t}\int\limits_0 t {g_o (t')}dt' $
$=$ $g_0 [1-(\frac{t}{\tau _{sw} })^{-1}(1-\exp (-t/\tau _{sw} ))]$.
If we produce this excitation using a $\pi $/2 pulse (i.e., ${g}'_0 (\tau )\tau 
=\pi /2)$ and measure the population of state $\vert $1$>$ after the 
excitation terminates (at $t=\tau$ ), we get a signal,

\begin{equation}
\vert C_1({g}'_0 (\tau ),\phi )\vert ^2$=1/2+$\eta $sin(2($\omega \tau +\phi )). 
\end{equation}

 This signal contains information of both the amplitude and the phase of the 
field \textbf{\textit{B}}$(t)$. The second term of Eq. 3 is related to the 
Bloch-Siegert shift [7,8] and we have called it the Bloch-Siegert 
oscillation (BSO) [2,3]. It is attributable to an interference between the 
so-called co- and counter-rotating parts of the oscillating field, with the 
atom acting as the non-linear mixer. For $\eta=0$, we have the 
conventional Rabi flopping that is obtained with the RWA. For a stronger 
coupling field, where the RWA is not valid, the second term of Eq. 3 becomes 
important [2,3], and the population will depend now both on the Rabi 
frequency and the phase of the driving field. In recent years, this effect 
has also been observed indirectly using ultra-short optical 
pulses [9,10,11] under the name of carrier-wave Rabi flopping. However, to 
the best of our knowledge, the experiment we report here represents the 
first direct, real-time observation of the absolute phase of an oscillating 
field using this self-interference mechanism. 

 From the oscillation observed, one can infer the value of ($\omega \tau 
+\phi $), \textit{modulo} $\pi $[12]. While the above analytical model 
presented here is  based on a two level system, practical examples of 
which are presented in reference 2, the effect is more generic, 
and is present even in three-  or  multi-level systems. In particular, 
we employed a three level system to observe this effect, due  
primarily to practical considerations. The specific 
system used consists of  three equally spaced Zeeman sublevels of 
$^{87}$Rb (5$^{2}$S$_{1/2}$: F=1: m$_{F}$=-1,0,1, denoted as states 
$\vert $0$>$, $\vert $1$>$ and $\vert $2$>$, respectively), where the 
degeneracy can be lifted by applying an external bias field. We have performed 
numerical simulations to confirm the presence of the BSO signature in the 
population dynamics of such a system as described below.

 Consider an equally spaced, ladder-type three-level system ($\left| 0 \right\rangle
$, $\left| 1 \right\rangle$ and $\left| 2 \right\rangle $ ). The transition
frequencies for $\left| 0 \right\rangle -\left| 1 \right\rangle $ 
and $\left| 1 \right\rangle -\left| 2 \right\rangle $ are of the same
magnitude $\varepsilon $. We also consider that a direct transition between
$ \left| 0 \right\rangle $ and  $\left| 2 \right\rangle $ is not allowed. 
Now, let the system be pumped by the same field at a frequency
$\omega $. Consider also that the Rabi frequency for the $\left| 0
\right\rangle -\left| 1 \right\rangle $ transition is $g_0 $ and that for 
$\left| 1 \right\rangle -\left| 2 \right\rangle $ is also $g_0 $. Then, the
Hamiltonian of the three-level system in a rotating frame can be written as:

\begin{equation}
\label{eq4}
\hat{\tilde {{H}'}}=-g_0 [1+\exp (-i2\omega t-i2\phi )]\;\,(\left| 0 
\right\rangle
\left\langle 1 \right|+\left| 1 \right\rangle \left\langle 2 \right|)+ c.c.
\end{equation}

\noindent where $\omega =\varepsilon $. The amplitudes of the three levels are
calculated numerically by solving the Schr\"{o}dinger equation for the above
Hamiltonian. The BSO amplitudes are then calculated by subtracting the   
population amplitude of each level \textit{with} the rotating wave approximation 
(RWA) from the population amplitude \textit{without }the RWA. The BSO oscillations 
for all the levels of such a system are shown in Fig. 1. 

The experimental configuration, illustrated schematically in Fig. 2, uses 
a thermal, effusive atomic beam. The RF field is applied to the atoms by a 
coil, and the interaction time $\tau $ is set by the time of flight of the 
individual atoms in the RF field before they are probed by a strongly 
focused and circularly polarized laser beam. The RF field couples the 
sublevels with $\vert \Delta $m$\vert $=1, as detailed in the inset of 
Fig. 2. Optical pumping is employed to reduce the populations of states 
$\vert $1$>$ and $\vert $2$>$ compared to that of state $\vert $0$>$ prior to the interaction with the 
microwave field.

A given atom interacts with the RF field for a duration 
$\tau$  prior to excitation by the probe  beam that couples state 
$\vert $0$>$ to an excited sublevel in 5$^{2}$P$_{3/2}$. 
The RF field was tuned to 0.5 MHz, 
with a power of about 10 W, corresponding to a Rabi frequency of about 4 MHz 
for the $\vert $0$>\rightarrow \vert $1$>$ as well as the $\vert $1$>\rightarrow 
\vert $2$>$ transition. The probe power was 0.5 mW focused to a spot 
of about 30 $\mu$m diameter, giving a Rabi frequency of about 60 
$\Gamma $, where $\Gamma (6.06 MHz)$ is the lifetime of the optical 
transition. The average atomic speed is 500 
m/s, so that the effective pulse width of the probe, $\tau _{LP}$, is 
about 60 nsec, which satisfies the constraint that $\tau 
_{LP}<<$1/$\omega $. Note that the resolution of the phase measurement 
is essentially given by the ratio of min [$\tau _{LP}$, $\Gamma ^{-1}$] 
and $1/\omega $, and can be increased further 

by making the  probe zone shorter. The fluorescence observed under 
this condition is essentially proportional to the population of level 
$\vert $0$>$, integrated over a duration of $\tau _{LP}$, 
which corresponds to less than 0.3 Rabi 
period of the RF driving field (for $g_{0M}/(2\pi) = 4$  MHz). 
Within a Rabi oscillation cycle, the BSO signal is maximum for 
$g_{0}(\tau)\tau/2 =(2n+1)\pi/2$, where n = 0, 1, 2,{\ldots}, 
so that there is at least one maximum of the BSO signal 
within the region of the probe.

Note that atoms with different velocities have different interaction times 
with the RF field, and produce a spread in the BSO signal amplitude within 
the probe region. However, \textit{the phase of the BSO signal is the 
same for all the atoms}, since it corresponds to the value of ($\omega 
\tau +\phi )$ at the time and location of interaction. Thus, there is no 
wash-out of the BSO signal due to the velocity distribution in the atomic 
beam.

Fig. 3 shows the spectrum of the observed BSO signal. In Fig 3(a), we show 
that the BSO stays mainly at 2$\omega $. When the probe beam is blocked, 
there is no signal (Fig.3(b)). When the RF intensity is increased a 
component of the BSO at 4$\omega $ begins to develop, as predicted. For 
the data in figure 4, the second harmonic of the driving field is used to 
trigger a 100 MHz digital oscilloscope and the fluorescence signal is 
averaged 256 times. When the probe beam is tuned to the 
$F=1\leftrightarrow F'=0$ transition, the population at $m=-1$ 
state is probed. When the probe is tuned to $F=1\leftrightarrow F'=1$, the 
combined populations of $m=-1$ and $m=0$  states are probed. That results in an 
effective detection of the complement of the population of $m=1$. On the other 
hand, when the probe beam is locked to the $F=1 \leftrightarrow F'=2$ 
transition, all three Zeeman sublevels of  $F=1$ are simultaneously probed and 
the phase information is not clearly present, since the total population of 
level $F=1$ is a constant. We observed  that the BSO signal amplitude varies as 
a function of an external magnetic  field applied in the $\hat{z}$ direction, 
with a peak corresponding to a  Zeeman splitting matching the applied frequency 
of 0.5 MHz. 

In figure 5, we show that the fluorescence signal is phase locked to the 
second harmonic of the driving field. First, we placed a delay line of 0.4 
$\mu$s on the cable of the reference field used to trigger the oscilloscope 
and recorded the fluorescence (Fig. 5a). Then, we put the 0.4$\mu $s delay 
line on the BSO signal cable, and recorded the fluorescence (Fig. 5b). The 
phase difference between the signals recorded in Figs. 5a and 5b is 
approximately 0.8 $\mu $s, as expected for a phase locked fluorescence 
signal. The data presented were for the probe resonant with the transition 
$F=1 \leftrightarrow F'=1$, but the same results were observed for 
$F=1 \leftrightarrow F'=0$.

To summarize, we report the first direct observation of the absolute phase 
of an oscillating electromagnetic field using self-interference in an atomic 
resonance.  This process is important in the precision of quantum bit rotations 
at a high speed.
The knowledge of the absolute phase of a RF field at a 
particular point of space may also be useful for single-atom quantum 
optics experiments. For example, an extension of this concept may possibly 
be used to teleport the wavelength of an oscillator, given the 
presence of degenerate distant entanglement, even in the presence 
of unknown fluctuations in the intervening medium [4-6,13]. Finally, 
this localized absolute phase detector
may prove useful in mapping of radio-frequency  fields in micro-circuits.  
Although a particular alkali atom was used in the present experiment, 
the mechanism is robust, and could be observed in virtually any atomic 
or molecular species.

Acknowledgements: We wish to acknowledge support from DARPA under the QUIST 
program, ARO under the MURI program, and NRO. 

\begin{figure}[htbp]
\centerline{\includegraphics[width=14cm,height=14cm]{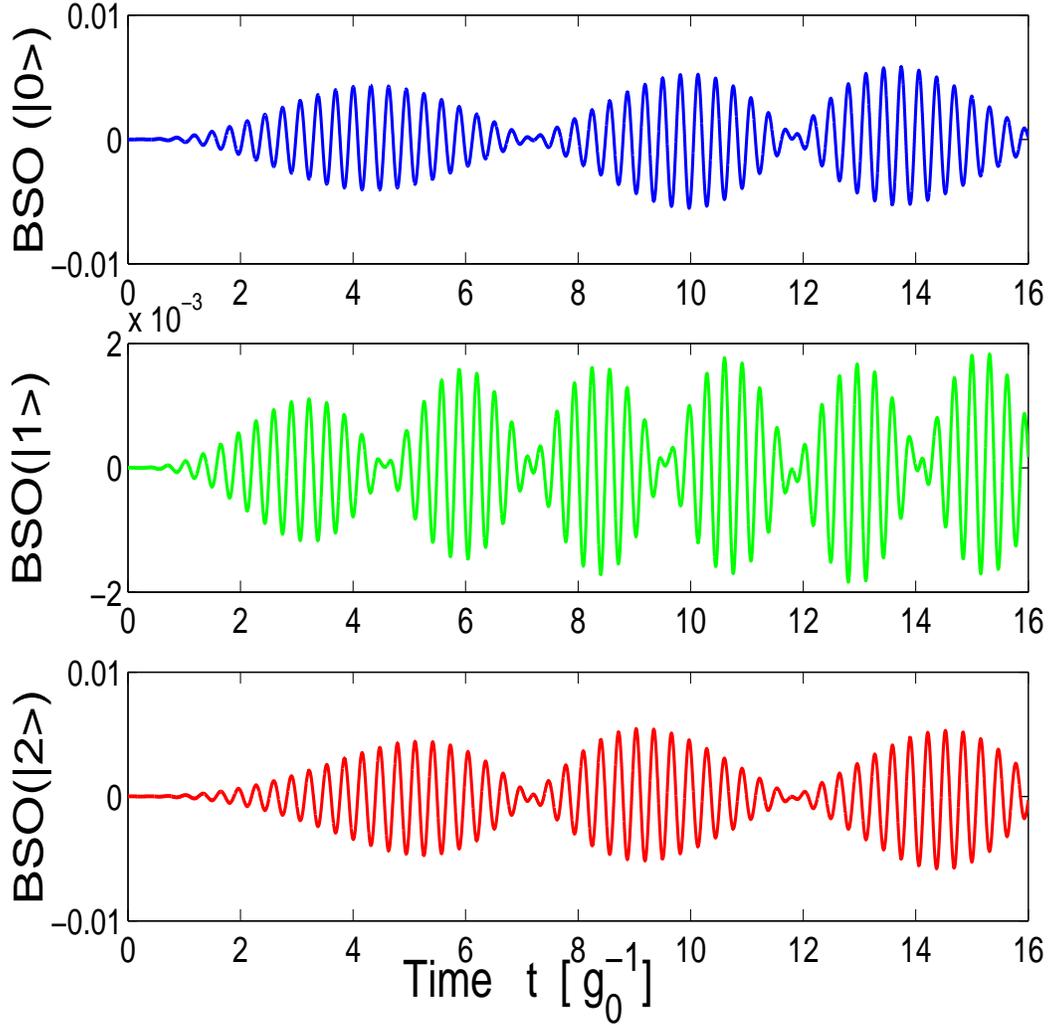}}
\caption{ BSO amplitude versus time t ( in units of
g$_{0}^{-1 })$ plots for all the levels of a three-level system. The  
initial densities of the levels are $\rho _{00} (t=0)=.5$, $\rho _{11}
(t=0)=.3$, and $\rho _{22} (t=0)=.2$, the Rabi frequency $g_0 =1$, and the
resonant frequencies $\omega _{02} =\omega _{21} =10$.}
\end{figure}

\begin{figure}[htbp]
\centerline{\includegraphics[width=18cm,height=15cm]{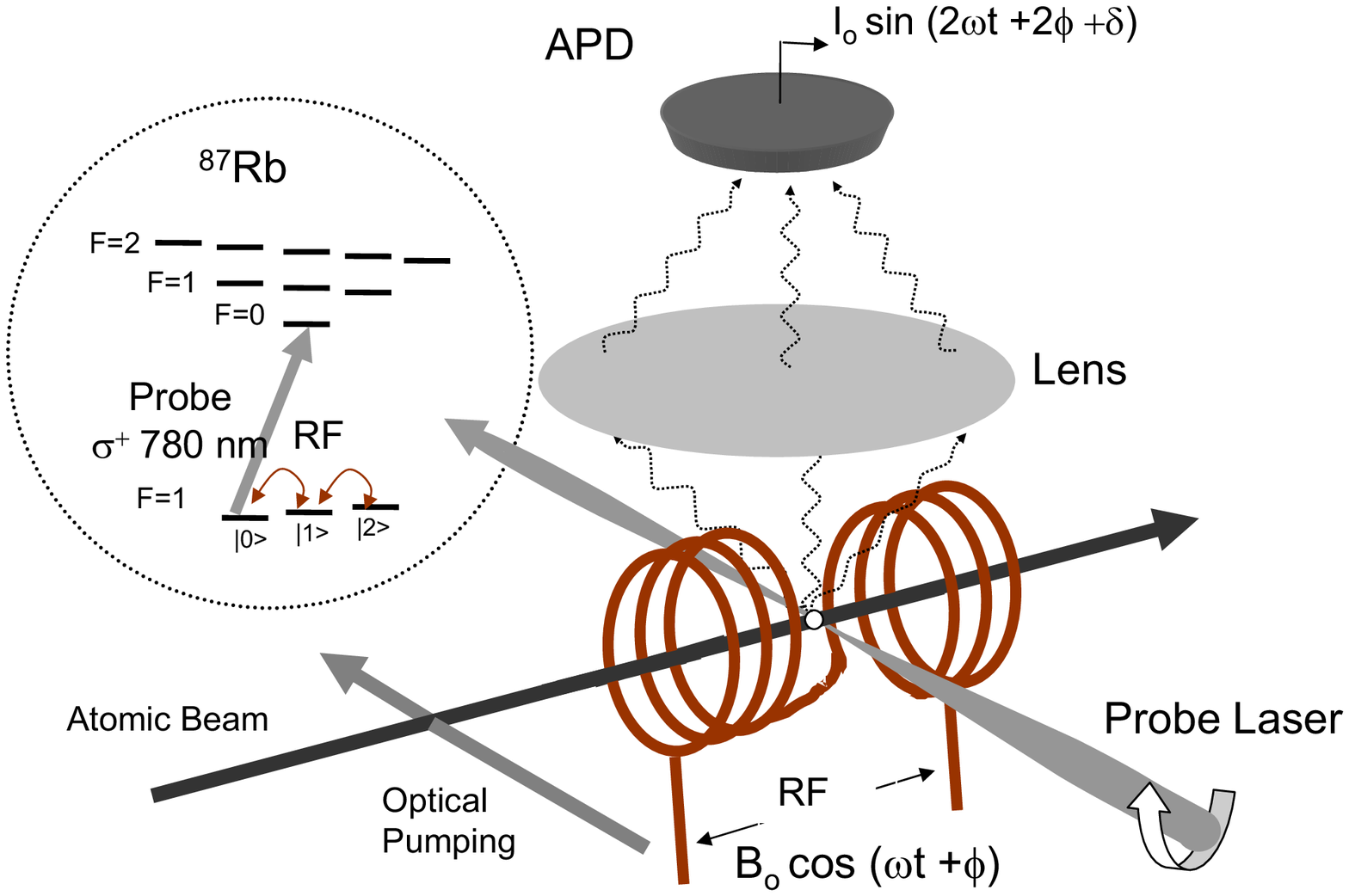}}
\caption{ experimental set-up. The 1 mm cross section 
rubidium atomic beam passes through the symmetry axis of the RF coil whose 
magnetic field is along the beam. The RF field of frequency \textit{$\omega $} 
is fed by a power amplifier connected to the resonant coil. 
A circularly polarized probe laser  beam is focused down to 30 $\mu $m in 
diameter through a gap in the middle  of the RF coil and perpendicularly to the 
atomic beam. The atomic fluorescence is collected by the lens and 
detected by an avalanche  photodiode (APD). The phase signature appears in the 
fluorescence signal  encoded in an oscillation at a frequency $2\omega$ due to 
the Bloch-Siegert  oscillations. In the picture, $\delta $  is an additional 
phase delay due to the APD circuits and cabling. 
\textbf{Inset:} Diagram of the relevant sublevels of 
the $D_{2}$ line of $^{87}Rb$. The numbers on the left represent the total 
angular momentum of the respective levels. The strong driving RF field couples to 
the ground state Zeeman sublevels. The probe beam must be resonant with an 
appropriate optical transition for the observation of the phase-locked 
signal, as discussed in the main text.} 
\end{figure}

\begin{figure}[htbp]
\centerline{\includegraphics[width=14cm,height=14cm]{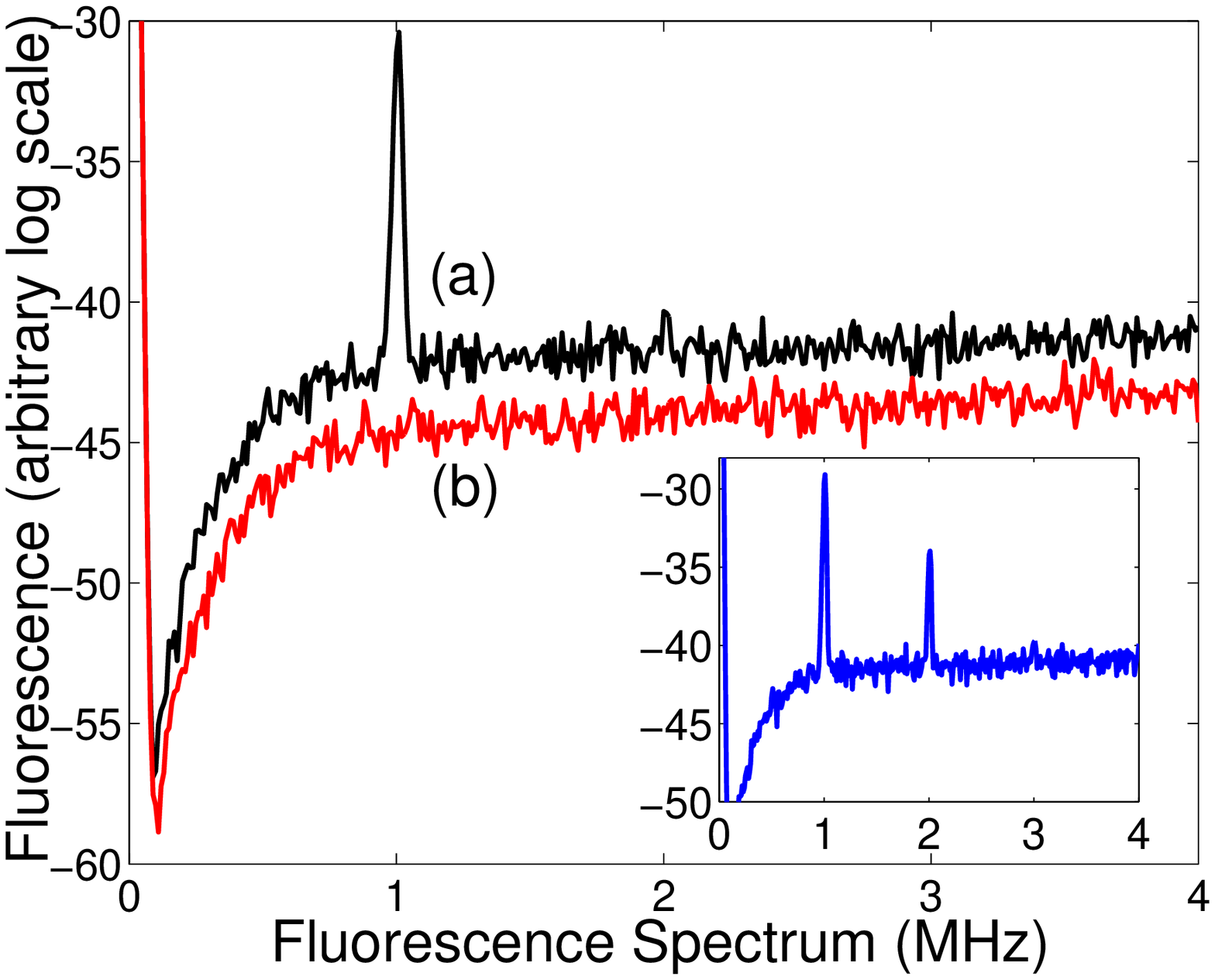}}
\caption{ Bloch-Siegert oscillation spectra. RF at 0.5 MHz and Rabi 
frequency around 4 MHz. (a) Probe beam resonant with the $5S_{1/2}$, $F=1$ 
$\leftrightarrow$  $5S_{3/2}$, $F'=0$ transition. The signal appears at $1 MHz$ 
with a linewidth less than 1 kHz (resolution limited by the spectrum 
analyzer). (b) Probe beam blocked. The dip structure around 100 kHz is an 
artifact due to the amplifier gain curve. \textbf{Inset}: Spectrum for same 
configuration and RF Rabi frequency around 10 MHz. Notice the 2 MHz harmonic 
which corresponds to the higher order BSO at $4\omega $.}
\end{figure}

\begin{figure}[htbp]
\centerline{\includegraphics[width=14cm,height=14cm]{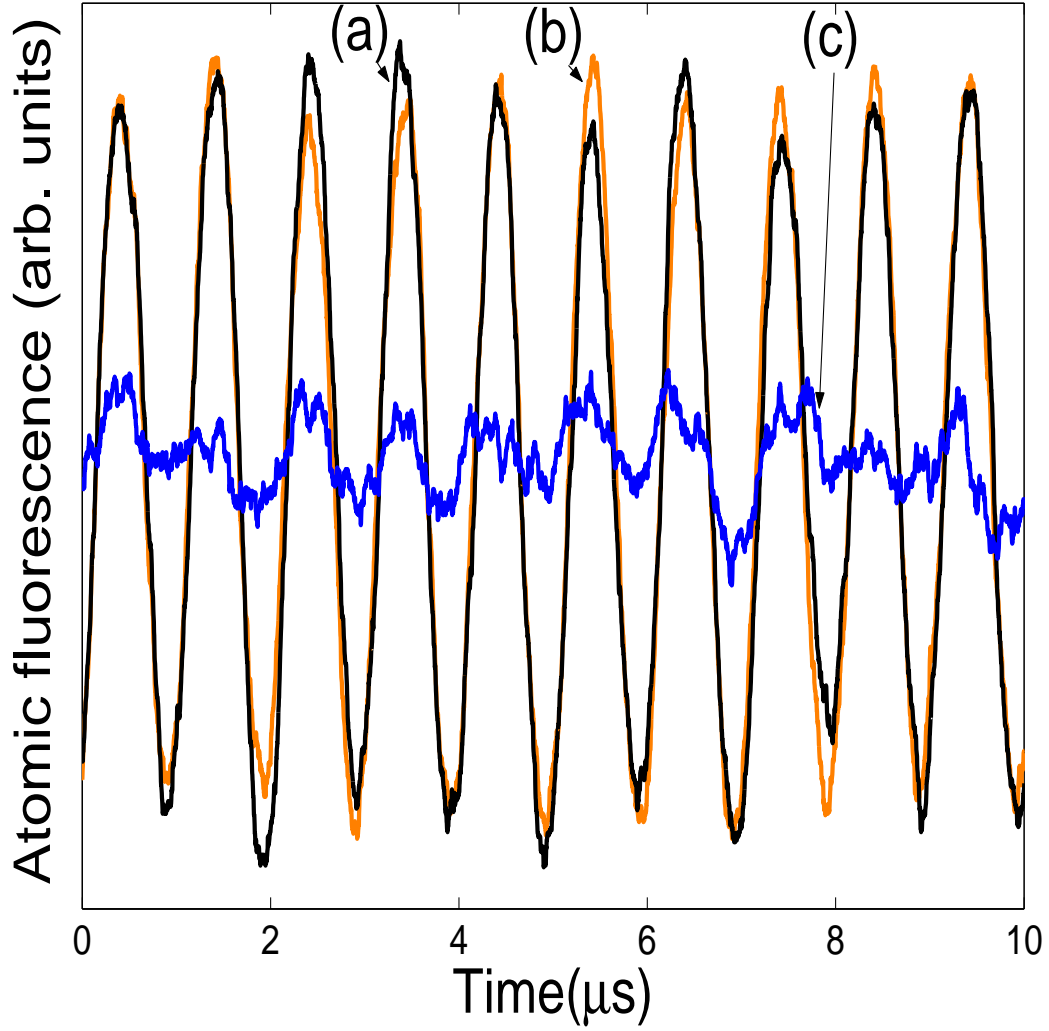}}
\caption{Time-dependence of the fluorescence signal at \textit{2$\omega $} when the 
probe beam is resonant to different excited states. The lines (a), (b), 
and the noisy  line (c) correspond to the probe locked to the transitions 
$F=1 \rightarrow F'=0$, $F=1 \rightarrow F'=1 $  and $F=1 \rightarrow F'=2$, 
respectively, of the $5S_{1/2} \rightarrow 5P_{3/2}$ transition in $^{87}Rb$ .}
\end{figure}

\begin{figure}[htbp]
\centerline{\includegraphics[width=14cm,height=14cm]{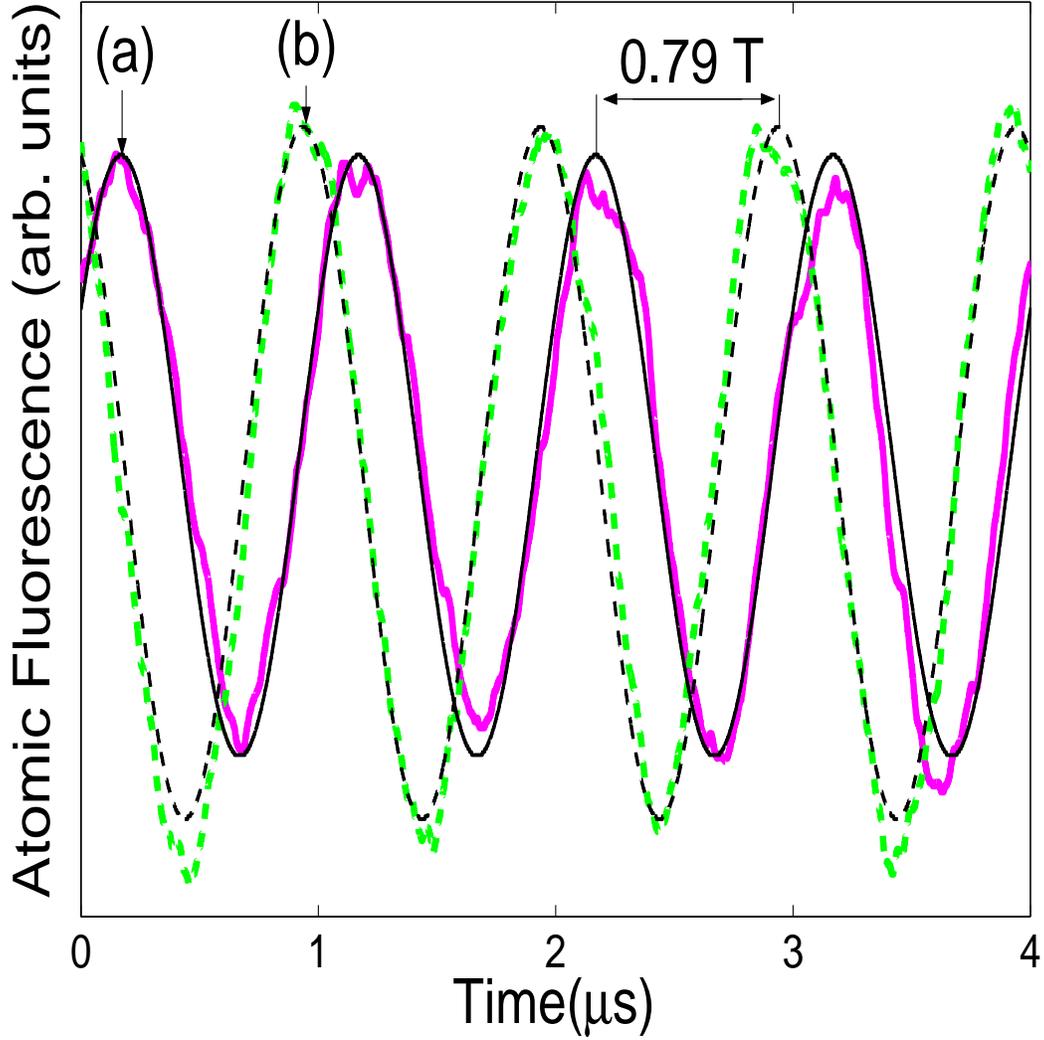}}
\caption{Demonstration of phase-locked fluorescence. T is the 
period of the Bloch-Siegert oscillation. (a) Population vs time when a 0.4T 
delay line was inserted in the reference field cable. (b) Population vs time 
when the same 0.4T  delay line was placed in the fluorescence signal 
cable. The figure shows that signal (b) is about 0.8T ahead of signal (a), 
confirming that the atomic fluorescence carries phase information which is 
locked to the absolute RF field phase. The solid and the dashed  
sinusoidal smooth curves are fittings to the experimental data, and was 
used for period and delay determination.}
\end{figure}


\begin{thebibliography}{20}
\bibitem{12} In principle, a simple modification of the experiment can be used to eliminate the modulus
           $\pi$ uncertainty. Specifically, if one applies a dc magnetic field parallel 
           to the RF field,  it leads to a new oscillation ( in the population of either level ) 
           at the fundamental frequency, with exactly the same phase as that of the driving field.
\end{thebibliography}
\end{document}